\documentclass[cits]{PoS}

\title{Application of Quadrature Methods for Re-Weighting in Lattice QCD}

\ShortTitle{Application of Quadrature Methods for Re-Weighting in Lattice QCD}

\author{\speaker{Abdou Abdel-Rehim}
\thanks{Current address: Computation-based Science and Technology Research Center (CaSToRC),
             The Cyprus Institute, 15 Kypranoros Str., P.O. Box 27456, 1645 Nicosia, Cyprus}\\
             Department of Physics, The College of William \& Mary   Williamsburg, VA 23187-8795, U.S.A\\
             E-mail: \email{a.abdel-rehim@cyi.ac.cy}}

\author{William Detmold \\
             Department of Physics, The College of William \& Mary   Williamsburg, VA 23187-8795, U.S.A \\
             and Jefferson Laboratory, 12000 Jefferson Avenue, Newport News, VA 23606, USA\\
             E-mail: \email{wdetmold@wm.edu}}

\author{Kostas Orginos\\
             Department of Physics, The College of William \& Mary   Williamsburg, VA 23187-8795, U.S.A \\
             and Jefferson Laboratory, 12000 Jefferson Avenue, Newport News, VA 23606, USA\\
             E-mail: \email{kostas@wm.edu }}

\abstract{Re-weighting is a useful tool that has been employed in Lattice QCD in different contexts
including, tuning the strange quark mass, approaching the light quark mass regime, and
simulating electromagnetic fields on top of QCD gauge configurations. In case of re-weighting
the sea quark mass, the re-weighting factor is given by the ratio of the determinants of two Dirac
operators $D_a$ and $D_b$. A popular approach for computing this ratio is to use a
pseudofermion representation of the determinant of the composite operator 
$\Omega=D_a(D_b^\dagger D_b)^{-1} D_a^\dagger$. Here, we study using quadrature methods 
together with noise vectors to compute the ratio of determinants. We show that, with quadrature 
methods each determinant can be computed separately using the operators $\Omega_a=D_a^\dagger D_a$ 
and $\Omega_b=D_b^\dagger D_b$. 
We also discuss using bootstrap re-sampling to remove the bias from the determinant estimator.
}

\FullConference{XXIX International Symposium on Lattice Field Theory \\
                          July $10-16$, 2011\\
                         Squaw Valley, Lake Tahoe, California}

\begin{document}

\section{Introduction}
Re-weighting is a useful tool that has been applied recently in lattice QCD simulations in different contexts. 
The main idea is that the expectation value of an observable $O$ with respect to an action $S_b$ can be written in terms 
of expectation values with respect to another action $S_a$ as follows:
\begin{eqnarray} 
\langle O \rangle_b & := & \frac{1}{Z_b}\int {\cal{D}} U \ O[U] \ e^{-S_b} \nonumber \\
                                &  =   & \frac{Z_a}{Z_b} \frac{1}{Z_a} \int {\cal{D}} U \ O[U] \ w[U] e^{-S_a} \nonumber \\ 
                                &   =   & \frac{ \langle O \ w \rangle_a} {\langle w \rangle_a },
\label{eq:reweighting_main_relation}
\end{eqnarray}
where $Z_a$ and $Z_b$ are the corresponding partition functions, $U$ is the gauge configuration and $w[U]$ is
called the reweighting factor corresponding to the gauge configuration $U$ and is defined by
$w[U]  :=  \frac{ e^{-S_b[U]} } { e^{-S_a[U]} }$. 
The labels "a" and "b" refers to a particular choice of the action parameters such as the sea quark mass.
Equation \ref{eq:reweighting_main_relation} is exact. However, when using importance sampling and for a
finite set of gauge configurations, reweighting could be biased. Because of this bias, reweighting is only
reliable when there is a small change between the actions $S_a$ and $S_b$, and the reweighting factors normalized by the ensemble mean are $O(1)$.
 
An important application of reweighting methods is {\it mass reweighting} in which 
observables corresponding to a sea quark mass $m_b$ are computed using configurations
generated with a sea quark mass $m_a$.
In this case the reweighting factor is given by
\begin{equation}
w[U] = \frac{ \det[(D(U;m_b)^\dagger D(U;m_b))^{\frac{n_f}{2}}]} {\det[(D(U;m_a)^\dagger D(U;m_a))^{\frac{n_f}{2}}]},  
\end{equation}  
where $D(U;m)$ is the Dirac operator for gauge configuration $U$ and bare quark mass $m$ and $n_f$ is the number of 
degenerate flavors being re-weighted. Applications of mass reweighting include tuning the strange quark mass in 
dynamical Domain-Wall simulations to its physical value \cite{Aoki:2010dy}, computing the strange Nucleon sigma 
term using the Feynman-Hellman theorem \cite{Ohki:2009mt}, simulating the $\epsilon$ regime \cite{Hasenfratz:2008fg}, 
and tuning to the physical point in 2+1 simulations of the PACS-CS collaboration \cite{Aoki:2009ix}. 

A common approach for computing the ratio of determinants is using pseudofermions. This has been the main 
technique used for mass reweighting \cite{Hasenfratz:2008fg,Aoki:2010dy,Aoki:2009ix}.
In this approach, the relation  
\begin{equation}
\frac{1}{\det(\Omega)} =  \frac{\int {\cal{D}}\xi^\dagger {\cal{D}} \xi \ e^{-\xi^\dagger \Omega \xi}} 
{\int {\cal{D}}\xi^\dagger {\cal{D}} \xi \ e^{-\xi^\dagger  \xi}}
\label{eq:psf_det}
\end{equation} 
is used to compute the determinant of a matrix $\Omega$. The right-hand side of Equation (\ref{eq:psf_det}) is computed
stochastically as an average over random gaussian fields $\xi$ generated with the distribution $e^{-\xi^\dagger \xi}$ as
\begin{equation}
< \frac{1}{\det(\Omega)} >_\xi =  < e^{-\xi^\dagger (\Omega-1) \xi} >_\xi, 
\label{eq:psf_det_stochastic}
\end{equation} 
where the notation $<g>_\xi$ means expectation value of $g$ over $\xi$. For a reliable estimate of $\det(\Omega)$ 
using Equation (\ref{eq:psf_det_stochastic}), $\Omega$ need to be close to the identity matrix. 
For $n_f$ degenerate flavors, 
\begin{equation}
\Omega=[D_a (D_b^\dagger D_b)^{-1} D_a^\dagger]^\frac{n_f}{2}.
\label{eq:compound_op}
\end{equation}

For small changes in the fermion action, $\Omega$ will be close to the identity matrix.
In addition, it is possible to
improve the estimation of $\det(\Omega)$ by dividing the change of the Dirac operator into smaller  steps and 
writing the ratio of determinants as a product of $k+1$ factors 
\begin{equation}
\frac{\det(D_a)}{\det(D_b)}= \frac{\det(D_a)}{\det(D_1)} \ \frac{\det(D_1)}{\det(D_2)} \ \dots  \frac{\det(D_{k-1})}{\det(D_k)} \ \frac{\det(D_k)}{\det(D_b)},
\label{eq:det_breakup}
\end{equation}
such that the ratio of determinants in each factor is close to one. The pseudofermion approach has the advantage of giving
directly an {\it unbiased} estimator of the ratio of determinants. It however requires the solution of a linear system
for each $\xi$. Another difficulty with the pseudofermion approach is that, for
a single flavor reweighting, it requires using a rational approximation for the square root of a matrix which in turn
requires a multi-shift solver. We study here an alternative way of computing the ratio of determinants based on the relation
$\det(\Omega) = e^{\rm{trace}[\log(\Omega)]}$.
In our approach, the trace of $\log(\Omega)$ is evaluated using noise methods and the matrix elements of $\log(\Omega)$ between
noise vectors are computed using Gauss quadratures \cite{Bai_somelarge}. This approach
has been proposed in the context of generating dynamical electromagnetic field configurations on top of existing QCD 
configurations \cite{Duncan:2004ys}. In this paper, we elaborate on using this approach also for mass
reweighting. In addition, we show that the need to solve a linear system for each noise vector can be avoided by computing the 
determinant of each Dirac operator separately and then take the ratio. This leads to a faster evaluation of the reweighting factor
than what is usually used in the literature. Another advantage of using the relation $\det(\Omega) = e^{\rm{trace}[\log(\Omega)]}$
is that including a fractional
power is trivial since
\begin{equation}
 \det(\Gamma^\alpha) = e^{\rm{trace}[\log(\Gamma^\alpha)]}= e^{\alpha\rm{trace}[\log(\Gamma)]}.
\label{eq:log_det_eq_tr_log}
\end{equation}
Finally, variance reduction techniques such as breaking the ratio of determinants into factors and dilution can be used.
It is noted that dilution couldn't be used with pseudofermions. 

\section{Quadrature and Noise approach}
Let $z$ be a vector whose elements are random variables satisfying
\begin{equation}
<z_i z_j>=<z_i^* z_j>=\delta_{ij}.
\end{equation}
Such vectors are called noise vectors. For a matrix $H$, we have
\begin{eqnarray}
\langle z^\dagger H z \rangle_{noise} & = & trace[H], \nonumber \\               
var(z^\dagger H  z)_{noise} & = & \sum_{i \neq j} |H_{ij}|^2+H_{ij} H_{ji}^*,
\end{eqnarray}
where $\langle \dots \rangle_{noise}$ and $var(\dots)_{noise}$ means the expectation value
and variance over the set of noise vectors. For a positive definite Hermitian matrix $H$, the element 
$z^\dagger H  z$ can be written as a Riemann-Stieltjes integral which can then be computed using 
Gauss quadrature methods (see \cite{Bai_somelarge} for more details).
For computing the ratio of determinants,  one could define $\Omega=D_a(D_b^\dagger D_b)^{-1}D_a^\dagger$ and
take $H:=log(\Omega)$, then use Equation [\ref{eq:log_det_eq_tr_log}]. In this case, $\Omega$ is close to the 
identity, however, each application of $\Omega$ to a vector requires the solution of a linear system which is 
expensive. We call this {\it the standard method}. Alternatively, one could compute each determinant separately
using the operators $\Omega_a=D_a^\dagger D_a$ and $\Omega_b=D_b^\dagger D_b$. Although these operators
are not close to the identity, they don't involve the inverse making this approach potentially faster. We call this
{\it the difference method}.

Our tests are done using Clover fermions. In this case it is advantageous to use 
use even-odd preconditioning. The Dirac operator is written in the form 
\begin{equation}
D = \left( \begin{array}{cc}
D_{ee} & D_{eo} \\
D_{oe} & D_{oo} \end{array} \right)
=\left( \begin{array}{cc}
  1 & 0 \\
D_{oe} D^{-1}_{ee} & 1
\end{array}
\right)
\left( \begin{array}{cc}
G   &  0 \\
0  &   Q \\
\end{array} \right)
\left( \begin{array}{cc}
1 & D^{-1}_{ee} D_{eo} \\
0 & 1 \\
\end{array} \right),
\end{equation}
where $G=D_{ee}$ and $Q = D_{oo} - D_{oe} D^{-1}_{ee} D_{eo}$ is the Schur complement.
The determinant of the Dirac operator is then given by $\det \left(D\right) = \det \left( G \right) \det \left( Q\right)$.
For this decomposition, $\det \left( G \right)$ can be computed exactly and cheaply while
$\det \left( Q \right)$ is estimated using noisy estimators.

Noisy estimators give an unbiased estimate for the trace of the logarithm. In order to get an unbiased estimate 
of the determinant by taking the exponential of the trace of the logarithm, additional techniques are needed.
There are two possible ways to obtain such unbiased estimator. The first method is using a resampling 
technique such as bootstrap (or jackknife). The second method, which is applicable to functions given as a power series, is based on stochastic 
summation of the series \cite{Bhanot:1985iq}. In our tests, we used bootstrap to obtain an unbiased estimate. 
Let $x_1,x_2,\dots,x_N$ be a sample of size $N$ of measurements of a random variable $x$ giving a sample average
$\overline{x}=\frac{1}{N}\sum_{i=1}^{N} x_i$. To obtain an unbiased 
estimator of the function $g(\langle x \rangle)$ where $\langle x \rangle$ is the true mean we generate $N_B$ 
bootstrap samples $x_i^\eta$ from the original data where $i=1,2,\dots,N$ and $\eta=1,2,\dots,N_B$. Compute
\begin{equation}
   x^B_\eta   =  \frac{1}{N} \sum_{i=1}^{N} x_i^\eta, \quad  g^B_\eta  = g(x^B_\eta), \quad
   \overline{g^B}   =  \frac{1}{N_B} \sum_{\eta=1}^{N_B} g^B_\eta, \quad 
   \overline{(g^B)^2}  =  \frac{1}{N_B} \sum_{\eta=1}^{N_B} (g^B_\eta)^2.
\end{equation}
The bootstrap estimator for $g(\langle x \rangle)$ is unbiased up to ${\cal{O}}(\frac{1}{N^2})$ and its error is given by
\begin{equation}
g(\langle x \rangle) \approx 2 \ g(\overline{x}) - \overline{g^B}, \quad
\sigma_{g(\overline{x})}^{bootstrap}=\sqrt{\frac{N}{N-1}} \sqrt{ \overline{{(g^B)^2}} - \overline{g^B}^2}.
\end{equation}
The bootstrap approach is applicable to a general function $g$. In our case $g(x)=e^x$.

\section{Results}
The method is tested on $16^3 \times 48$ lattices with tadpole improved tree-level Symanzik gauge action at $\beta=6.5$
corresponding to lattice spacing $a\approx 0.09 \ fm$. 
A tadpole improved clover fermion action is used with one level of stout
smeared links with smearing parameter $\rho=0.125$. We have three degenerate flavors with quark masses roughly in 
the range of the strange quark mass.
Configurations were generated with bare quark mass parameters $m_q=-0.170,-0.175,-0.180$ with corresponding
lattice pion masses $am_\pi=0.3570(16), 0.3302(25), 0.3097(18)$ respectively. The corresponding pion masses in physical 
units are $m_\pi(MeV)=780(35), 720(55),680(39)$. In our tests, reweighting was used to compute observables for bare sea quark masses
$m_q=-0.175$, and $-0.180$ by reweighting configurations generated with $m_q=-0.170$. We have analyzed $700$ configurations.
The fact that we have configurations generated with $m_q=-0.175$ and $-0.180$ allowed us to check the correctness of the 
reweighting procedure by comparing to expectation values of observables computed on configurations generated with the correct 
sea quark mass. Calculations done using Chroma software \cite{Edwards:2004sx}.
In order to improve the accuracy of the computation of the matrix element $z^\dagger log(\Omega) z$, we use double precision 
for the quadrature Lanczos algorithm with single precision for the matrix-vector multiplication. We use complex ${\mathbb{Z}}_4$ noise
vectors where each element of $z$ can have the values $1,-1,i,-i$ with equal probability.

First, we compare  ${\rm trace}({\rm log}(Q_a^\dagger Q_a))-{\rm trace}({\rm log}(Q_b^\dagger Q_b))$ (difference method)
to \\
${\rm trace}({\rm log}[Q_a(Q_b^\dagger Q_b)^{-1}Q_a^\dagger])$ (standard method). Mathematically, the results should be identical and 
stochastically they should be consistent within errors. For this test, we use $1000$ ${\mathbb{Z}}_4$ noise vectors and $m_a=-0.170$ and $m_b=-0.180$.
To investigate the effect of numerical precision, we do the comparison for the situation where all the computation is done in double precision (both matrix-vector
multiplication, the linear solver, and the Lanczos quadrature algorithm is done in double precision), and when mixed precision
calculation is used (matrix-vector multiplication and the linear solver is done in single precision, while dot products
inside the Lanczos quadrature algorithm are done in double precision). The Conjugate Gradient (CG) algorithm is used to solve 
linear systems when the standard method is used. In the case of double precision calculation,
the tolerance for the linear system is set to $1e-10$ and the quadrature for computing the matrix element is also computed
to converge to the same tolerance. In the mixed precision calculation, the linear solver and the quadrature is set to converge 
to a tolerance of $1e-7$ when the standard method is applied. When the difference method is applied, the quadrature for each matrix 
element is set to $1e-9$. The reason it was necessary to set the tolerance of the quadrature in the case
of the difference method to $1e-9$ instead of $1e-7$ as it was in the standard case is to have the same number of 
significant digits (7 digits) after taking the difference. In Table \ref{table:comp_standard_diff_method_nodil_nobreak}, we compare the
results which show that we get results which are consistent with each other within statistical errors. This is important since the difference 
method is much faster than the standard method. In Table \ref{table:comp_stand_diff_matvecs_170_180}, we compare the cost of each method in terms of
 the number of matrix-vector
multiplication each method uses. In the case of the standard method, the Lanczos quadrature algorithm takes few iterations, about 5 iterations, however
each iteration involves the solution of a linear system using CG which takes about $500$ iterations (a matrix-vector
product here means multiplication of a vector with $Q^\dagger Q$). The quadrature algorithm converges in few iterations because the 
compound operator is close to the identity. On the other hand, for the difference method, no linear system need to be solved, however, the Lanczos 
quadrature algorithm takes more iterations to converge since the operators $Q_a^\dagger Q_a$ and $Q_b^\dagger Q_b$ are far from the identity.
The total cost of the difference method is much smaller than the standard method. 

Since both methods give the same results, we use the difference method for mass re-weighting. We reweight configurations generated with sea quark mass
$-0.170$ to compute observables corresponding to sea quark masses $-0.175$ and $-0.180$. We computed the reweighting factor for 
$700$ configurations and for the production calculations we use only $500$ noises and a mixed precision calculation in which the matrix-vector products
are done in single precision while dot products in the Lanczos quadrature algorithm are done in double precision. The matrix elements were made to converge
to tolerance $1e-9$.

\begin{figure}[htbp]
\begin{center}
\includegraphics[width=0.40\textwidth,keepaspectratio]{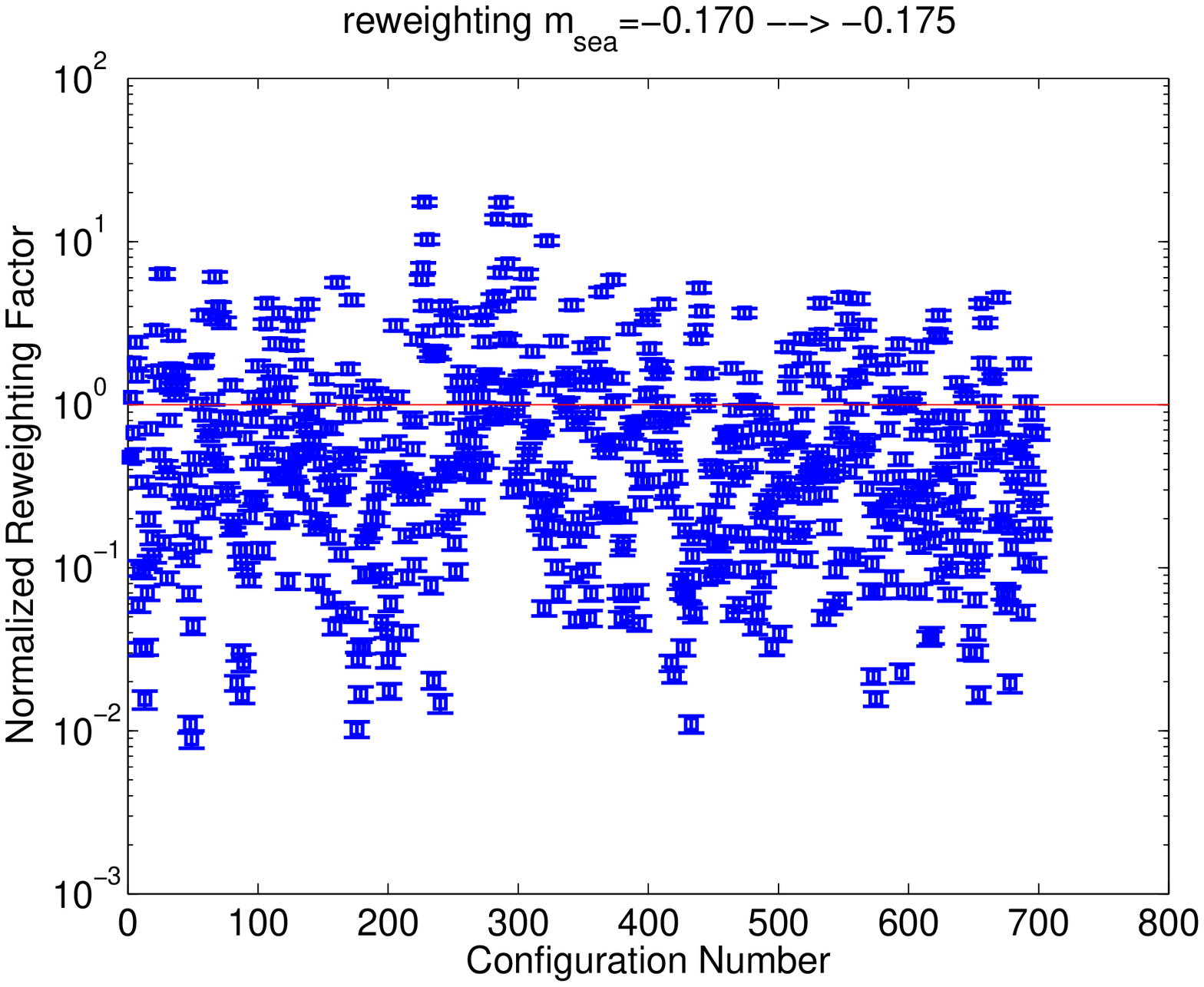}
\includegraphics[width=0.40\textwidth,keepaspectratio]{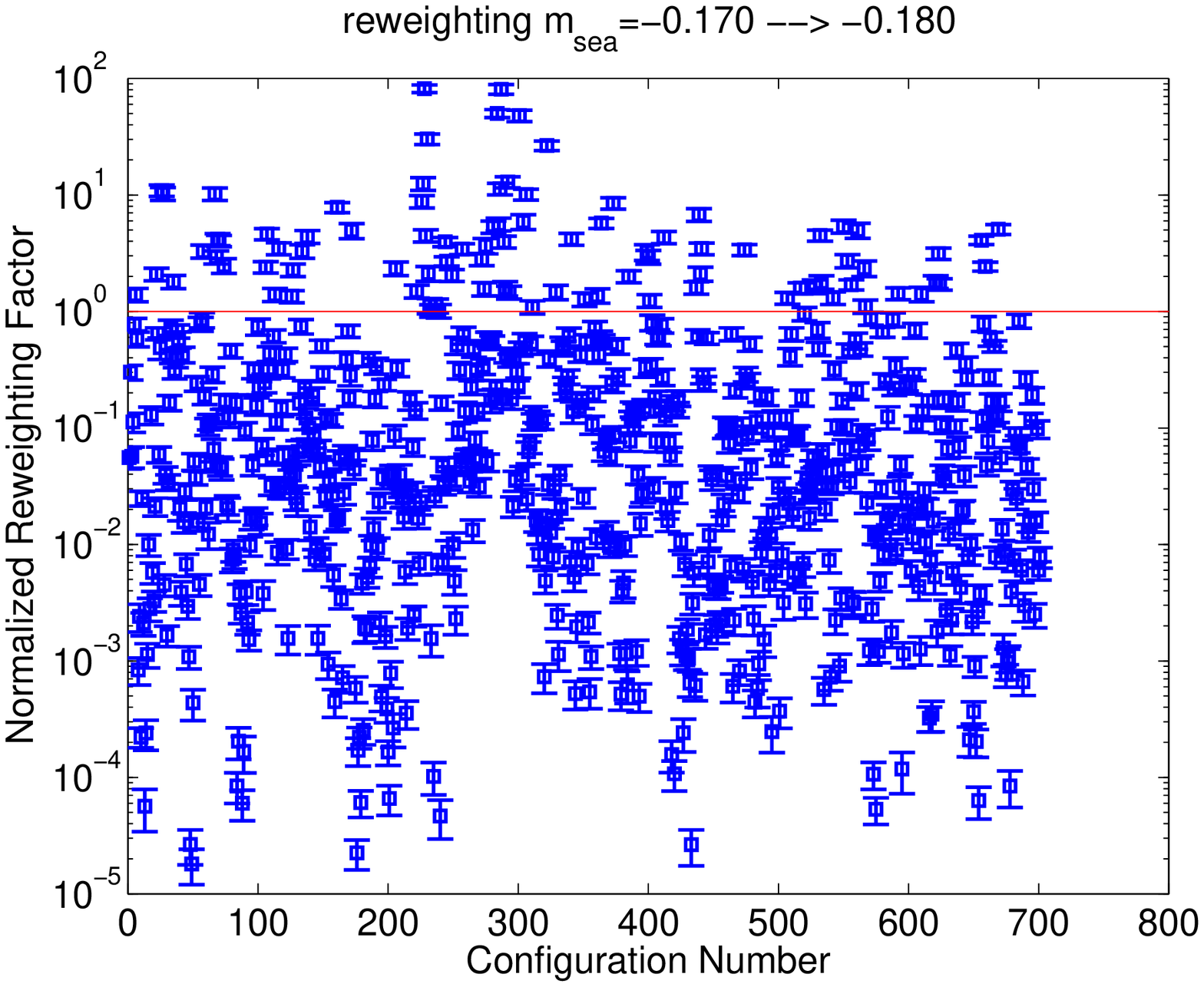}
\end{center}
\caption{The normalized re-weighting factors. Left: reweighting from sea quark mass $-0.170$ to $-0.175$.
Right: reweighting from sea quark mass $-0.170$ to $-0.180$.}
\label{fig:normalized_rw_factors}
\end{figure}

\begin{figure}[htbp]
\begin{center}
\includegraphics[width=0.40\textwidth,keepaspectratio]{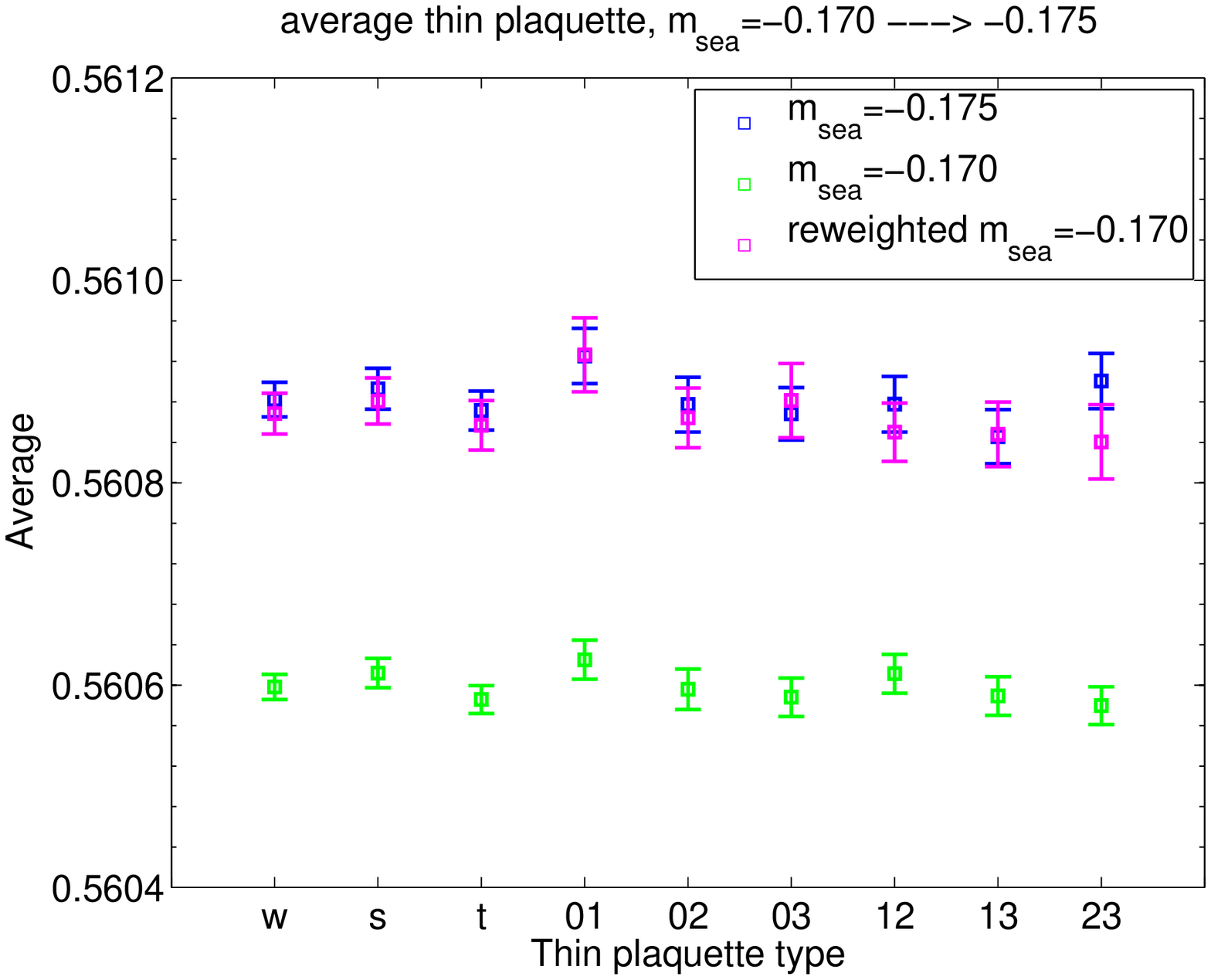}
\includegraphics[width=0.40\textwidth,keepaspectratio]{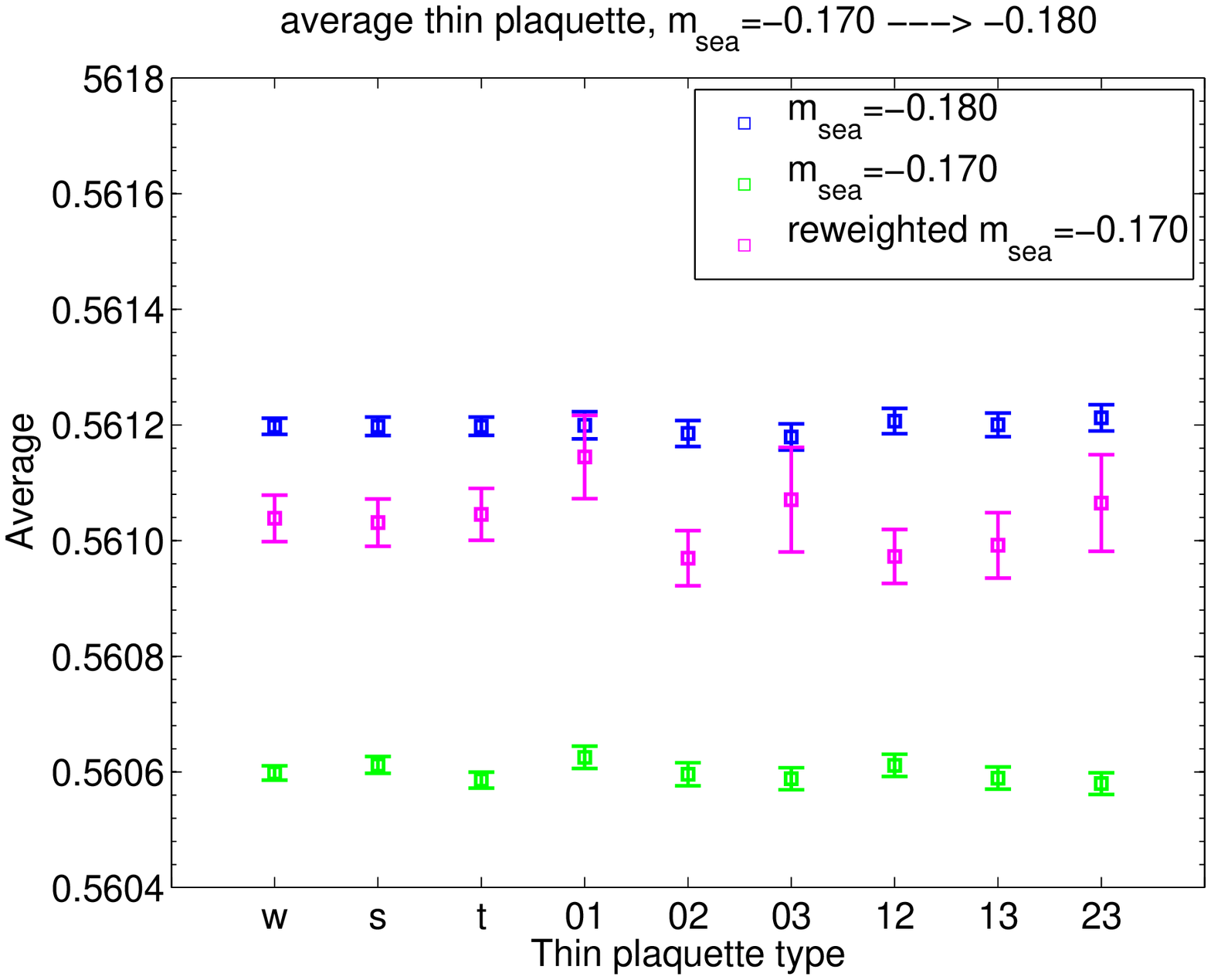}
\end{center}
\caption{Reweighted thin plaquettes compared with correct values. The horizontal axis corresponds to different types of thin plaquettes (different planes,
plaquettes in the spatial directions only, etc.). Left: reweighting from sea quark mass $-0.170$ to $-0.175$.
Right: reweighting from sea quark mass $-0.170$ to $-0.180$.}
\label{fig:comp_ave_thin_plaq}
\end{figure}

In Figure \ref{fig:normalized_rw_factors}, we show the normalized re-weighting factors. As shown in the figure, there is a larger fraction of the configurations
with small reweighting factor for the $-0.180$ case than the $-0.175$ case.  Having the re-weighting factor available, we can now re-weight any physical 
observable of interest. As an example, we look at re-weighting the average thin plaquette. In Figure \ref{fig:comp_ave_thin_plaq}, we compare re-weighted 
various thin plaquettes to the correct values obtained from configurations generated with the correct sea quark mass. The results show that $700$ configurations
provide enough statistics to re-weight from sea quark mass $-0.170$ to $-0.175$, but not to correctly reweight from $-0.170$ to $-0.180$.
\begin{table}[htbp]
\begin{tabular}{lcccc}
\hline
\hline
Config. No. & Standard Method    & Difference Method& Standard Method   & Difference Method\\
            & Double Precision   & Double Precision & Mixed Precision   & Mixed Precision  \\
\hline
1 & 5485.76(9)  & 5485.78(8) & 5485.91(9)  & 5485.77(8) \\ 
2 & 5484.34(9)  & 5484.35(8) & 5484.49(9)  & 5484.33(8) \\
3 &  5486.57(9) & 5486.63(8) & 5486.73(9)  & 5486.62(8) \\
\hline
\hline
\end{tabular}
\caption{Comparison of ${\rm{trace}}\{\log[Q_a(Q_b^\dagger Q_b)^{-1}Q_a^\dagger]\}$ (standard method) and 
${\rm{trace}}\{\log[Q_a^\dagger Q_a]\}-{\rm{trace}}\{\log[Q_b^\dagger Q_b]\}$ (difference method) on three 
configurations where $Q_a$ corresponds to $m_a=-0.170$ and $Q_b$ corresponds to $m_b=-0.180$ using $1000$ noises
without dilution.}
\label{table:comp_standard_diff_method_nodil_nobreak}
\end{table}

\begin{table}[htbp]
\begin{tabular}{lcccc}
\hline
\hline
Config. Num. & Standard Method  & Difference Method & Standard Method  & Difference Method \\
             & Mixed Precision  & Mixed Precision  & Double Precision & Double Precision  \\
\hline
1  &  5+2671  &  105+123   & 6+5164   &  137+162 \\
2  &  5+2652  &  103+121   & 6+5007  &   137+163 \\
3  &  5+2685  &  106+123   & 6+5172  &   138+165 \\
\hline
\hline
\end{tabular}
\caption{Comparison of the number of matrix-vector products used by
the standard and difference methods for a single noise vector when $S_a$ corresponds to 
$m_0=-0.170$ and $S_b$ corresponds to $m_0=-0.180$.
For the standard method, the first number is the number of Lanczos algorithm iterations and the second number is the
total number of iterations used by the CG solver during these iterations. For the difference 
method, the two numbers correspond to the number of iterations used by each Lanczos algorithm.}
\label{table:comp_stand_diff_matvecs_170_180}
\end{table}

\section{Conclusions}
The application of Lanczos quadrature algorithm in mass re-weighting is studied.
It is shown that it is possible to avoid the need to solve a linear system and apply 
the algorithm to compute each determinant separately before taking the ratio. 
We also showed how to get an un-biased estimator for the reweighting factors using bootstrap method.
Having configurations generated with the sea quark mass we are trying to re-weight to, allowed
us to compare the re-weighted observable to the correct ones and check for the range of applicability 
of the re-weighting procedure.

\subsection{Acknowledgements}
We would like to thank Stefan Meinel, Balint Joo, Robert Edwards, and Anna Hasenfratz for valuable discussions.
A. Abdel-Rehim, W. Detmold and K. Orginos were supported in part by DOE grants DE-AC05-06OR23177 (JSA) and DE- FG02-04ER41302. 
W. Detmold was also supported by DOE OJI grant DE-SC0001784 and Jeffress Memorial Trust, grant J-968. A. Abdel-Rehim would like
to thank the Cyprus Institute for support during the writing of this report.  


\begin{thebibliography}{99}
\bibitem{Ferrenberg:1988yz}
 A.~M.~Ferrenberg, and R~.H~.Swendsen, Phys.Rev.Lett. 61, 1988, 2635-2638.
\bibitem{Hasenfratz:2008fg}
A.~Hasenfratz, R.~Hoffmann, and S.~Schaefer, Phys. Rev. D78, 2008, 014515, [arXiv:0805.2369]
\bibitem{Aoki:2010dy}
Y. Aoki et. al., Phys.Rev. D83 (2011) 074508, [arXiv:1011.0892].
\bibitem{Ohki:2009mt}
H. Ohki et. al., pos{PoS(LAT2009)124}, [arXiv:0910.3271].
\bibitem{Aoki:2009ix}
S. Aoki, et. al., Phys.Rev. D81 (2010) 074503 [arXiv:0911.2561].
\bibitem{Bai_somelarge}
Z. Bai, M. Fahey, and G. Golub, J. Comput. Appl. Math. 74 (1994)71--89.
\bibitem{Duncan:2004ys}
A. Duncan, E. Eichten, and R. Sedgewick,
Phys.Rev.  D71 (2005) 094509 [hep-lat/0405014].
\bibitem{Bhanot:1985iq}
G. Bhanot and A.D. Kennedy, Phys.Lett. B157(1985)70.
\bibitem{Edwards:2004sx}
  R.~G.~Edwards and B.~Joo, 
  Nucl.\ Phys.\ Proc.\ Suppl.\  {\bf 140}, 832 (2005)
  [arXiv:hep-lat/0409003].

\end{thebibliography}
\end{document}